\documentstyle[aps,prl]{revtex}
\begin{document}
\input epsf.sty
\twocolumn[\hsize\textwidth\columnwidth\hsize\csname %
@twocolumnfalse\endcsname
\draft
\widetext

\title{Common features of nanoscale structural correlations in magnetoresistive
manganites with ferromagnetic low-temperature state}

\author{V. Kiryukhin$^1$, T. Y. Koo$^1$, A. Borissov$^1$,
Y. J. Kim$^2$, C. S. Nelson$^2$, 
J. P. Hill$^2$, D. Gibbs$^2$, and S-W. Cheong$^{1,3}$}
\address{(1) Department of Physics and Astronomy, Rutgers University,
Piscataway, New Jersey 08854}
\address{(2) Department of Physics, Brookhaven National Laboratory, Upton,
New York 11973}
\address{(3) Bell Laboratories, Lucent Technologies, Murray Hill,
New Jersey 07974}

\date{\today}
\maketitle

\begin{abstract}
We report x-ray scattering studies of nanoscale structural correlations 
in Nd$_{1-x}$Sr$_x$MnO$_3$ and La$_{1-x}$(Ca,Sr)$_x$MnO$_3$,
$x$=0.2--0.5. We find that the correlated 
regions possess a temperature-independent correlation length of 2-3
lattice constants which is the same in all samples.
The period of the lattice modulation of the correlated regions is 
proportional to the Ca/Sr doping concentration $x$. Remarkably, the
lattice modulation periods of these and several other manganites with
a ferromagnetic ground state fall on the same curve when plotted 
as a function of $x$.
Thus, the structure of the correlated regions in these materials appears
to be determined by a single parameter, $x$. We argue that these observations
provide important clues for understanding the Colossal Magnetoresistance
phenomenon in manganites.

\end{abstract}

\pacs{PACS numbers: 75.30.Vn, 71.38.+i, 71.30.+h}

\phantom{.}
]
\narrowtext

Manganite perovskites of the chemical formula A$_{1-x}$B$_x$MnO$_3$ (where
A is a rare earth, and B is an alkali earth atom) have recently attracted 
considerable attention because they exhibit a wide diversity of ground states
and a number of interesting phase transitions
\cite{Review}. Perhaps the most dramatic of these is the
magnetic-field-induced insulator-metal 
transition which is referred to as the Colossal
Magnetoresistance (CMR) effect. In its most widely studied form, the CMR
effect is the transition from a paramagnetic insulating (PI) to a ferromagnetic
metallic (FM) phase. The very large difference between the electrical 
resistivities of the PI and FM phases lies at the core of the CMR effect, and
considerable efforts have been spent in order to explain this difference
\cite{Review}. The 
metallic nature of the FM phase was explained in the 1950's in the
framework of the double-exchange (DE) mechanism \cite{Review} in which the
itinerant $e_g$ electrons have their spins aligned with the localized $t_{2g}$
spins of the Mn atoms by virtue of a strong Hund coupling.
The DE mechanism, however, does not explain the high resistivity of the PI
phase \cite{Millis}. While recent experimental \cite{Polarons,xrayPolarons}
and theoretical \cite{Millis} work suggests that the enhanced resistivity 
might in fact result from the presence of small lattice polarons, 
the anomalously large resistivity of the PI phase still
remains largely unexplained. 

With the resurgent interest in the manganites in the 1990's, it was realized
that the rich behavior of the manganites results from the complex interplay
between the charge, spin, lattice, and orbital degrees of freedom
\cite{Review}. 
A very important consequence of the competition between the 
various degrees of 
freedom is the large variety of inhomogeneous states exhibited by the 
manganites \cite{D}. The characteristic length scale of these inhomogeneous
states varies from 
microns in conventional phase-separated states down to nanometers
in the materials exhibiting nanoscale charge/orbitally ordered regions
\cite{Review,D,Martensit,Corr}. 

There is a growing body of evidence that the large resistivity of the PI
phase in manganites is associated with nanoscale inhomogeneities.
In fact, it was recently found that the PI phase
exhibits short-range structural correlations \cite{Corr,Corr1}.
It was proposed that these correlations reflect the presense of
nanoscale correlated regions possessing local
charge/orbital order \cite{Corr,Corr1}. The associated Jahn-Teller lattice
distortions form a periodic lattice modulation in the ordered regions. 
The electrical resistivity of the PI state has been shown
to increase with increasing concentration of these 
correlated regions \cite{Corr}, 
suggesting that the large resistivity of this state is due to the presence
of such regions. Therefore, understanding the properties of the correlated 
regions is a necessary step toward
understanding the transport properties of 
the PI phase and the mechanism of the CMR effect.

There is, however, only limited and indirect experimental information about
the local structure of the correlated nanoregions. 
In this work, we report a
systematic x-ray diffraction study of the structural correlations in
Nd$_{1-x}$Sr$_x$MnO$_3$, $x$=0.3--0.5, and La$_{1-x}$(Ca,Sr)$_x$MnO$_3$,
$x$=0.2, 0.25.
We find that the correlation length
of these regions is the same in all the samples studied and does not depend on
temperature. Further,
the period of the lattice modulation in the correlated regions was found to vary
with Sr/Ca doping
concentration $x$, following an approximately linear relationship
with $x$. Remarkably,
the lattice modulation
periods of the investigated samples, as well as of several other manganites with
a ferromagnetic ground state, fall on the same curve.
Thus, the structure of the correlated regions appears
to be common to all these materials, and 
determined by a single parameter, $x$. These observations impose
strict constraints on possible models of the structure of the correlated
domains as
well as on the mechanism for their formation and 
should, therefore,
provide important clues for understanding the CMR
phenomenon in manganites.  
 
Single crystals of Nd$_{1-x}$Sr$_x$MnO$_3$ ($x$=0.3, 0.45, and 0.5),
La$_{0.8}$Ca$_{0.2}$MnO$_3$, and 
La$_{0.75}$(Ca$_{0.45}$Sr$_{0.55}$)$_{0.25}$MnO$_3$ were grown using the
floating zone technique. X-ray diffraction measurements were carried out
at beamlines X22C and X20C at the National Synchrotron Light Source. 
A 10 keV x-ray beam was focused by a mirror, monochromatized by a 
double-crystal Ge (111)
monochromator, scattered from the sample,
and analyzed with a pyrolytic graphite crystal.
The samples were mounted
in a closed-cycle refrigerator (T=10-450 K). In this paper,
Bragg peaks are indexed in the orthorhombic {\it Pbnm}
notation in which the longest
lattice constant is $c$, and scattering vectors ($h, k, l$) are given in
reciprocal lattice units.

\begin{figure}
\centerline{\epsfxsize=2.9in\epsfbox{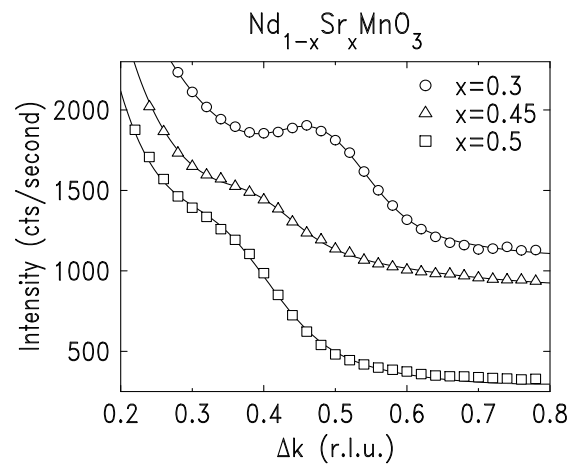}}
\vskip 5mm
\caption{X-ray scans along the (4, 6+$\Delta k$, 0) direction 
(Nd$_{0.55}$Sr$_{0.45}$MnO$_3$ and Nd$_{0.5}$Sr$_{0.5}$MnO$_3$),
and the (4, 4+$\Delta k$, 0) direction (Nd$_{0.7}$Sr$_{0.3}$MnO$_3$).
The temperatures are 210 K, 275 K, and 260 K for the $x$=0.3, 0.45, and 0.5
samples, respectively. The solid lines are the results of fits, as
discussed in the text. For clarity, $x$=0.3 and $x$=0.45 data are shifted up 
by 750 cts/second.}
\label{fig1}
\end{figure}

\begin{figure}
\centerline{\epsfxsize=2.9in\epsfbox{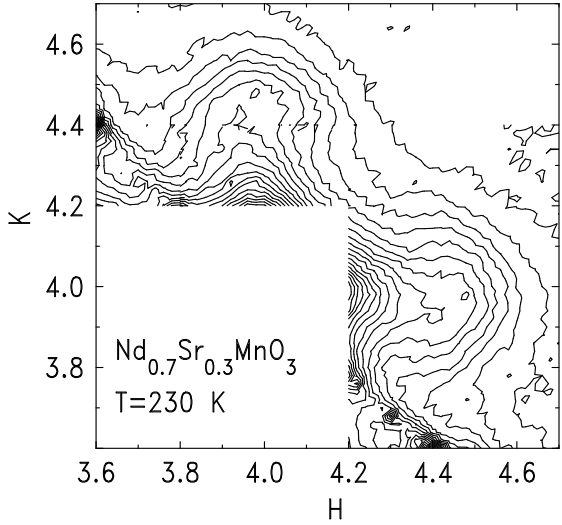}}
\vskip 5mm
\caption{Contour plot of the x-ray intensity around the
(4, 4, 0) Bragg peak at T=230 K in Nd$_{0.7}$Sr$_{0.3}$MnO$_3$.}
\label{fig2}
\end{figure}

We first focus our discussion on the properties of the Nd$_{1-x}$Sr$_x$MnO$_3$ 
samples with $x$=0.3--0.5. 
These are paramagnetic insulators at high temperatures.
With decreasing temperature, they undergo a transition to a ferromagnetic 
metallic state at T$_c$$\approx$210 K, 280 K, and 250 K for $x$=0.3, 0.45, 
and 0.5, respectively \cite{NSMO}. In addition, the $x$=0.5 sample
undergoes a transition to a CE-type 
charge-ordered state at
T$_{co}$=150 K. Fig. \ref{fig1} shows that
in the PI phase, all these samples exhibit broad peaks
in scans taken along the $b^*$ direction in reciprocal space.
These broad peaks arise due to the presence of correlated nanoregions
\cite{Corr}. 
The intensity of these peaks reflects
the concentration of the correlated regions \cite{Disclamer}, 
the peak width is inversely
proportional to their size (correlation length), and the peak position defines
the period of the lattice modulation in the correlated regions. 
The peaks are observed on top of a sloping background which is attributed to
scattering from uncorrelated polarons, also known as 
Huang scattering, and to thermal-diffuse
scattering \cite{xrayPolarons}.
The overall scattering pattern is illustrated in
Fig \ref{fig2}, which shows a contour plot of the x-ray intensity in the
$x$=0.3 sample at T=230 K around the (4, 4, 0) Bragg peak. 
In this figure, two broad peaks at (4.5, 4, 0) and 
(4, 4.5, 0) are superimposed on a ``butterfly-like'' shaped background which
is characteristic of single-polaron scattering in 
the manganites \cite{xrayPolarons}.
The broad peaks are observed along both the $a$ and $b$ crystallographic
directions. However, it is not possible to establish with any certainty
whether this is because there are two different modulation vectors present or
merely arises from crystallographic twinning in the sample.

The data of Fig. \ref{fig1}, and subsequent data taken as a function
of temperature, were fitted to a sum of a
Gaussian line shape and a monotonically sloping background, the latter 
described by a power-law function. 
Several other descriptions were also tried, but the resulting parameters were
not found to be significantly different. The error bars in the figures reflect
the total parameter variation, including that due to different possible fitting
functions.

\begin{figure}
\centerline{\epsfxsize=2.9in\epsfbox{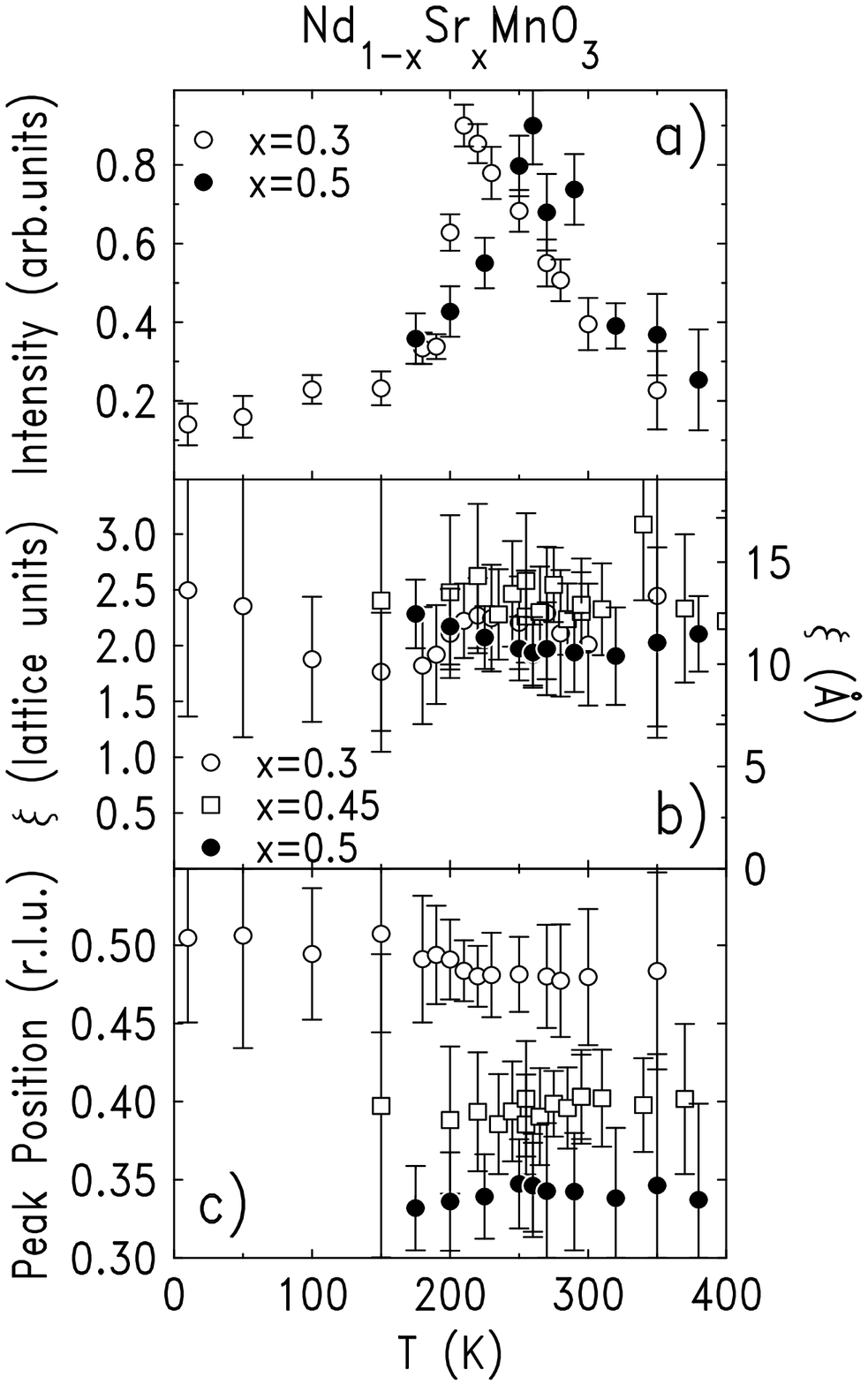}}
\vskip 5mm
\caption{(a) Temperature dependence of the intensity of the peak due to the
structural correlations in Nd$_{1-x}$Sr$_x$MnO$_3$. 
The single-polaron background is subtracted as
discussed in the text. (b) The correlation length of the ordered regions.
(c) The position of the peak relative to
the nearest Bragg peak (the lattice modulation wave vector).}
\label{fig3}
\end{figure}

Fig. \ref{fig3}(a) shows the intensity of the 
correlated peaks in the $x$=0.3 and 0.5 samples as a function of temperature
(the $x$=0.45 data are omitted for clarity). As previously found in 
La$_{0.7}$Ca$_{0.3}$MnO$_3$ \cite{Corr}, 
this intensity is strongly reduced in the
ferromagnetic metallic phase. However, the structural correlations do not
completely disappear below the Curie temperature, and in the $x$=0.3 sample
small concentrations of the correlated regions survive in the FM region of
the phase diagram even at the lowest temperatures.
This observation is
consistent with the results of neutron measurements of the pair-distribution
function in related samples \cite{PDF}. These latter
measurements showed that local
Jahn-Teller lattice
distortions, albeit in small concentrations,
are still present in the FM phase. Our data demonstrate
that at least some of these distortions stem from the correlated regions
discussed in this work.

The strong reduction of the correlations below T$_c$ is consistent with a
picture in which the $e_g$ electrons in the correlated regions 
are localized: as
these regions disappear, the resistivity decreases. 
While the correlations
are suppressed below T$_c$ in all our samples, the transition in the
$x$=0.3 sample is much sharper than that in the $x$=0.5 sample. This
observation is in general agreement with recent reports of anisotropy and 
possible inhomogeneity of the FM state in 
Nd$_{0.5}$Sr$_{0.5}$MnO$_3$ \cite{NSMO1}. 

\begin{figure}
\centerline{\epsfxsize=2.9in\epsfbox{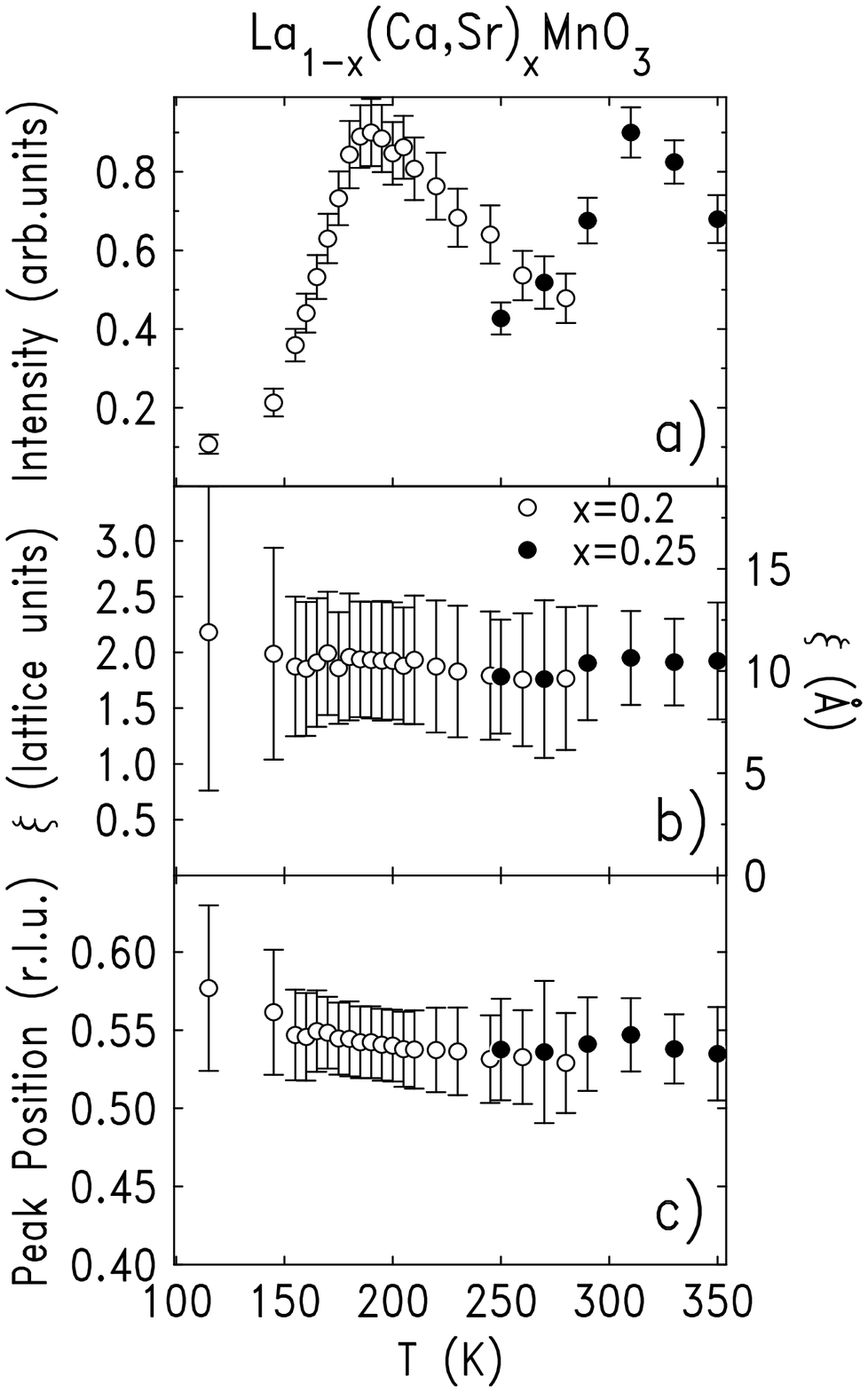}}
\vskip 5mm
\caption{(a) Temperature dependence of the intensity of the peak due to the
structural correlations in La$_{0.8}$Ca$_{0.2}$MnO$_3$, and
La$_{0.75}$(Ca$_{0.45}$Sr$_{0.55}$)$_{0.25}$MnO$_3$. 
The single-polaron background is subtracted as
discussed in the text. (b) The correlation length of the ordered regions.
(c) The position of the peak relative to
the nearest Bragg peak (the lattice modulation wave vector).}
\label{fig3.5}
\end{figure}

La$_{0.8}$Ca$_{0.2}$MnO$_3$ and
La$_{0.75}$(Ca$_{0.45}$Sr$_{0.55}$)$_{0.25}$MnO$_3$ samples with
T$_c\approx$190 K and 300 K, respectively, exhibited 
transport, magnetic, and structural 
properties similar to those of Nd$_{1-x}$Sr$_x$MnO$_3$. Nanoscale structural
correlations in these samples were analyzed in the same manner as those
of Nd$_{1-x}$Sr$_x$MnO$_3$. The results of this analysis are shown in 
Fig. \ref{fig3.5}. Figs. \ref{fig3} and \ref{fig3.5} show that the qualitative
features of the nanoscale correlations in the (Nd,Sr)MnO$_3$ and
(La,Ca,Sr)MnO$_3$ samples are very similar.

Figs. \ref{fig3}(b) and \ref{fig3.5}(b) show 
the correlation length of the ordered regions.
The correlation length is defined as the inverse half-width-at-half-maximum of
the diffraction peak. While this definition gives the correct correlation
length for exponentially decaying correlations and Lorentzian line shape, 
it does not necessarily
provide a good measure for the ``size'' of the ordered regions. For example,
simple calculations of the structure factor for perfectly ordered clusters of a
finite size show that in this case
the correlation length, as defined above, underestimates
the cluster size by as much as a factor of two.
Therefore, even though the
correlation length $\xi$ 
determined in our experiments is between 2 and 3 lattice
constants, the number of three-dimensional unit cells in the ordered regions 
could be as large as (2$\xi$)$^3$, {\it i.e.} about one hundred.

The data of Figs. \ref{fig3}(b), \ref{fig3.5}(b)
show that the size of the correlated regions 
does not change with temperature and is the same for all the samples, to within
errors. This same size was also previously obtained for correlated regions in
La$_{0.7}$Ca$_{0.3}$MnO$_3$ samples \cite{Corr1}.
Taken together, these results suggest
that there is a common mechanism 
defining the size of the correlated regions in these materials. 

\begin{figure}
\centerline{\epsfxsize=2.9in\epsfbox{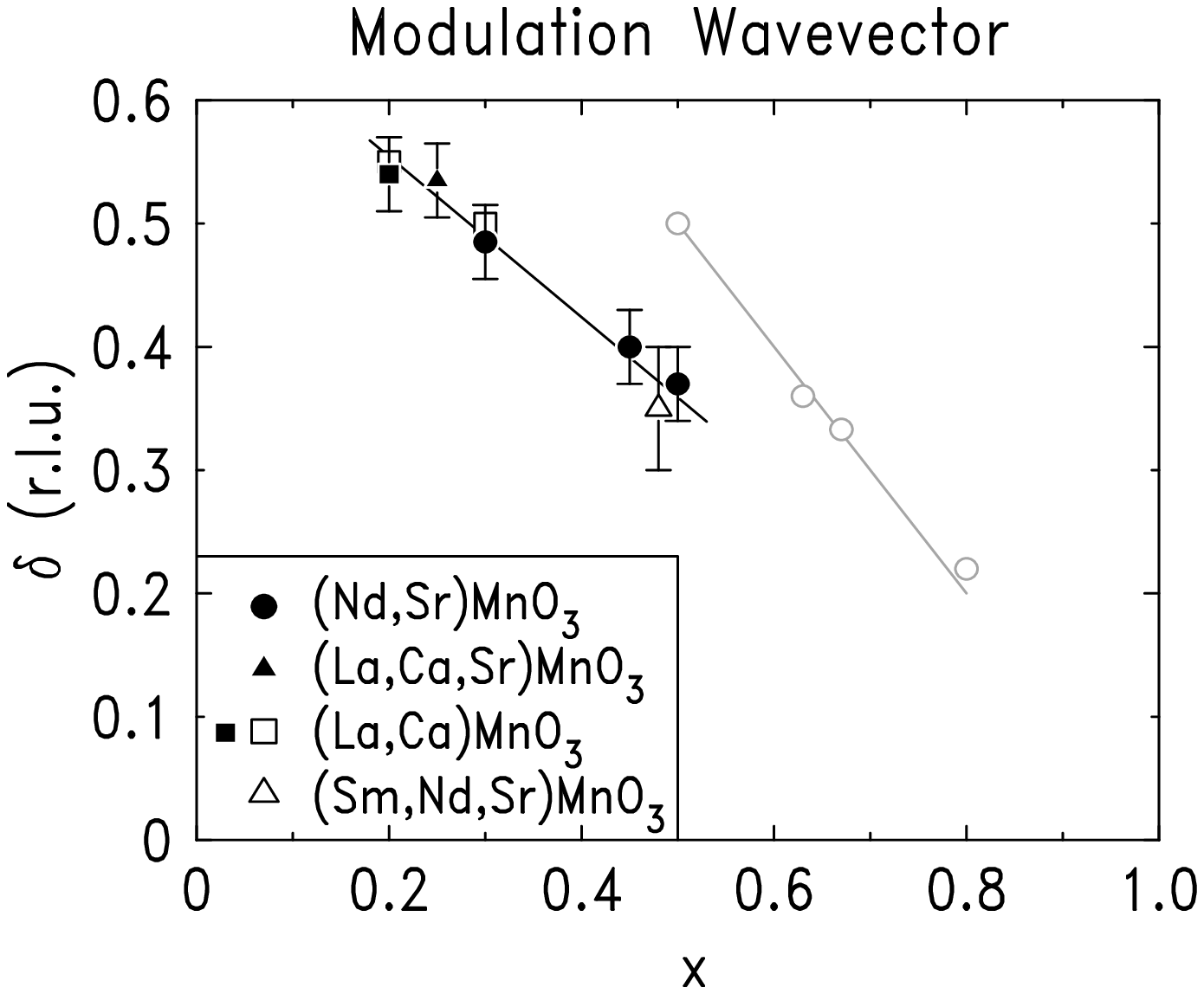}}
\vskip 5mm
\caption{The wave vector $\delta$
of the lattice modulation in the correlated regions as a
function of doping $x$ (black symbols). 
Filled symbols represent our own data. 
The data shown with open symbols were taken from Refs. [4,7,8].
Grey symbols show the wave vector of the structures with
{\it long-range} charge/orbital order
observed in manganites with $x>$0.5 (Ref. [18]).}

\label{fig4}
\end{figure}

The period of the lattice modulation in the correlated regions 
is shown in Figs. \ref{fig3}(c), \ref{fig3.5}(c). This period is different in
different samples. However, as was the case for the correlation length, it is
temperature-independent, to within the experimental errors. 
In Fig. \ref{fig4}, we plot the period of the
lattice modulation as a function of doping, $x$. We also show
results of previous measurements in La$_{0.8}$Ca$_{0.2}$MnO$_3$,
La$_{0.7}$Ca$_{0.3}$MnO$_3$, and 
(Sm$_{0.875}$Nd$_{0.125}$)$_{0.52}$Sr$_{0.48}$MnO$_3$ samples from Refs. 
\cite{Corr}, \cite{Corr1}, and \cite{xrayPolarons} which all have FM
ground states.
Remarkably, all the results shown in Fig. \ref{fig4}
appear to fall on the same curve. This observation, together with the
independence of the nanoregion size on temperature and sample composition,
strongly suggests that the structure of the correlated regions in
these compounds is defined by a single parameter, $x$.

These observations impose constraints on any possible models for
the structure of the correlated regions, as well as on the
mechanism responsible for their formation. 
In particular, they indicate that the mechanism controlling
the period of the lattice modulation in these regions acts
via the Ca/Sr doping concentration $x$ in the sample. 
In the first approximation, the carrier concentration in the samples is
proportional to $x$, and therefore the carrier concentration might, in fact,
be the parameter defining the structure of the correlated regions.
In addition,
there appears to be a common mechanism
defining the size of the correlated regions in these materials.
While such a mechanism is currently unknown, 
we would like to make several comments on the 
possible structure of the ordered regions. First, because of the characteristic
lattice distortion and the insulating properties of the correlated regions, 
it is
likely that they possess charge and orbital order. Moreover, for
$x$=0.3, the lattice modulation has the same wave vector as the CE-type
charge/orbitally ordered state. For $x$=0.3, therefore, 
it is reasonable to assume
that the correlated regions
are in fact small domains possessing the CE-type order
with its checker-board-type charge order and the characteristic
orbital ordering \cite{Review}.
As $x$ grows, different charge/orbitally ordered structures, possibly containing
discommensurations \cite{Review}, may be realized. Finally,
as $x$ approaches 0.5, the period of the lattice modulation 
approaches 3, and one of the ``striped'' structures observed in highly
doped-manganites \cite{Review}
could be realized. As discussed above, the small correlation lengths observed in
our experiments are compatible with all these structures.
 
If the above scenario is correct, one has to explain why the proposed
charge-ordered structures are observed at doping levels different from 
those of the
corresponding long-range ordered counterparts. The CE-type order, for example,
is most stable at $x$=0.5, and the striped structure with a period of 3 is
observed at $x$=2/3 (see grey symbols in Fig. \ref{fig4}). 
One possible reason for this discrepancy might be the
nanoscale phase separation in which charge-rich and charge-poor regions are 
formed. It is possible that the charge concentration in the charge-depleted
correlated regions corresponds to the ideal concentration required for the
formation of the charge-ordered structures with the observed periodicity. 
Such variations in the charge concentration could result from nanoscale
chemical inhomogeneities, such as clustering of the cations in the 
A-position. This scenario would naturally explain the independence on
temperature of the correlation length and the modulation wavevector in the
correlated regions, but does not provide any straightforward explanation for the
observed dependence of the modulation wavevector on the sample composition.

Even in the absence of chemical inhomogeneities, however, formation of
charge-rich and charge-poor regions is still possible. 
A number of theoretical calculations, in fact, predict such a phase
separation \cite{D}, and some experimental evidence has recently become
available \cite{Alonso}. 
Note, this nanoscale phase separation is different from the well-known case
of phase separation in the manganites, in which domains with sub-micrometer
size are formed \cite{D}. In the latter case, Coulomb 
forces prevent any significant charge redistribution. 
The Coulomb forces should not, however, prevent the formation
of nanoscale charge-depleted regions. They can, nevertheless, be one of the
factors limiting the region size.
Of course, other mechanisms explaining the experimental observations are 
possible. For example, nesting properties of the Fermi surface 
have been proposed for the origin
of the structural correlations observed in two-dimensional 
manganites \cite{Chu}. However, preliminary calculations suggest that
this scenario is unlikely in the
three-dimensional manganites discussed here \cite{Millis2}.
In another recent work \cite{Last}, effects of quenched disorder are 
considered, but detailed predictions about the structure of the PI state are
yet to be achieved.
Clearly, further work is needed
to establish the actual mechanism leading to the formation of the correlated
regions.

Finally, we would like to emphasize that all of the samples considered above 
exhibit the FM phase at low temperatures (the exception is 
Nd$_{0.5}$Sr$_{0.5}$MnO$_3$, in which 
the lowest-temperature charge-ordered state
is separated from the PI phase by the intermediate-temperature FM state).
The case of the transition from the PI or FM 
phase directly to the charge-ordered
phase, such as in Pr$_{1-x}$Ca$_x$MnO$_3$ or La$_{0.5}$Ca$_{0.5}$MnO$_3$,
is more complex. In such transitions, the periodicity of
the lattice modulation and the correlation length vary on warming as
the low-temperature, long-range charge order disappears and the 
high-temperature correlations arise \cite{Corr1,PCMO}. 
The existing experimental data 
are not completely consistent, and
it is unclear as to what extent they reflect the 
intrinsic properties of the PI state. Systematic measurements at temperatures
much larger than the charge-ordering transition temperature are needed to
answer this question.

In summary, we report that the nanoscale structural correlations 
characteristic of 
the paramagnetic state of magnetoresistive manganites possess a common
correlation length and that the structure of the correlated regions
appears to be determined by a single parameter
-- the concentration of the doped divalent ions. We argue that observation of
such common features imposes strict constraints on the possible
mechanism responsible for the formation of the 
correlated regions, and therefore
should provide important clues for understanding the CMR mechanism in
manganites.  

We are grateful to A. J. Millis for important 
discussions. This work was supported by the NSF under grants No. 
DMR-0093143, DMR-9802513, by the DOE under contract No.
AC02-98CH10886, and by the NSF MRSEC program, Grant No. DMR-0080008.


\begin{references}
\bibitem{Disclamer} This intensity can also reflect the amplitude of the
lattice modulation in the correlated regions. In principle, this can change 
with temperature. However, this 
scenario appears to be inconsistent with the results of the neutron
measurements of Ref. [11]
\end{references}
\end{document}